\documentclass[aps,preprint,showpacs]{revtex4}
\usepackage{amssymb}

\usepackage{graphicx}

\begin{document}

\title{Big Bounce Singularity of a Simple Five-Dimensional Cosmological
Model }
\author{Lixin Xu}
\author{Hongya Liu}
\thanks{hyliu@dlut.edu.cn}
\author{Beili Wang}
\affiliation{Department of Physics, Dalian University of Technology, Dalian,116024, P.R.
China}

\begin{abstract}
The big bounce singularity of a simple 5D cosmological model is studied.
Contrary to the standard big bang space-time singularity, this big bounce
singularity is found to be an event horizon at which the scale factor and
the mass density of the universe are finite, while the pressure undergoes a
sudden transition from negative infinity to positive infinity. By using
coordinate transformation it is also shown that before the bounce the
universe contracts deflationary. According to the proper-time, the universe
may have existed for an infinitely long time.
\end{abstract}

\maketitle

The inflationary cosmology can resolve three important problems of the
standard big bang models: the galaxy formation problem, the horizon problem,
and the flatness problem. However, there are other deep questions of
cosmology which inflation does not resolve \cite{1}: What occurred at the
initial singularity? Does time exists before the big bang? These issues have
been popular in cosmology for a long time. Tolman \cite{2} firstly discussed
an oscillating cosmological model within the framework of general
relativity. He pointed out that the main difficulty of such an oscillating
model is that the universe has to pass through a cosmological singularity on
each bounce, and during each cycle, enormous inhomogeneities would
undoubtedly be generated. This is the so-called entropy problem of the
oscillating models. Recently, an ekpyrotic cosmological model was presented
by Khoury \textit{et. al.} within the framework of the brane world scenario 
\cite{3,4}. According to this model, our big bang universe emerges
from a collision between two branes. When the two branes collide
inelastically and bounce off one another, brane kinetic energy is partially
converted into matter and radiation and our universe begins to expand. In
the ekpyrotic model the universe undergoes a single transition from
contraction to expansion \cite{3,4}. Drawn from his idea, Steinhardt
and Turok presented a cyclic model in which the universe undergoes an
endless sequence of cosmic cycles each of which begins with a ``big bang''
and ends with a ``big crunch'' \cite{1}. It was argued that in both the
ekpyrotic and cyclic model all major cosmological problems may be resolved
without any use of inflation \cite{1,3,4}.

In this letter, we will discuss the big bounce cosmological model presented
recently by Liu and Wesson \cite{5}. This model differs from Tolman's
oscillating model as well as the cyclic model in that the universe in the
big bounce model undergoes a single transition from contraction to
expansion. It also differs from the ekpyrotic model in that the big bounce
universe contracts, before the bounce, deflationary from an empty de Sitter
vacuum \cite{5}. We will focus on a simple exact solution of the big bounce
model and study what occurred at and before the bounce.

The idea of extra spatial dimensions comes from the attempt of unifying
gravity with other interactions. The space-time-matter (STM) theory
developed by Wesson and coworkers is inspired by the unification of matter
and geometry \cite{6,7}. In this theory, our 4D space-time is embedded
in a 5D Ricci-flat manifold, and the 4D matter fields are assumed to be
``induced'' from pure geometry in 5D. Mathematically, the STM theory
strongly relies on Compbell's theorem which states that any solution of
N-dimensional Einstein equations can locally be embedded in a Ricci-flat
(N+1)-dimensional manifold \cite{8}. It has been show that the STM theory
agrees with all the classical tests of general relativity in the solar
system \cite{9}, and it also gives physically interesting effects such as a
new (fifth) force \cite{10}. There are equivalence between STM and brane
model \cite{11}. Recently, the bounce cosmology has been used to construct
brane models \cite{12}.

Within the framework of the STM theory, an exact 5D cosmological solution
was given by Liu and Mashhoon in 1995 \cite{13}. Then, in 2001, Liu and
Wesson restudied the solution and showed that it describes a cosmological
model with a big bounce as opposed to a big bang \cite{5}. The 5D metric of
this solution reads

\begin{equation}
dS^{2}=B^{2}dt^{2}-A^{2}\left( \frac{dr^{2}}{1-kr^{2}}+r^{2}d\Omega
^{2}\right) -dy^{2}  \label{5-metric}
\end{equation}
where $d\Omega ^{2}\equiv \left( d\theta ^{2}+\sin ^{2}\theta d\phi
^{2}\right) $ and

\begin{eqnarray}
B &=&\frac{1}{\mu }\frac{\partial A}{\partial t}\equiv \frac{\dot{A}}{\mu } 
\nonumber \\
A^{2} &=&\left( \mu ^{2}+k\right) y^{2}+2\nu y+\frac{\nu ^{2}+K}{\mu ^{2}+k}.
\label{A-B}
\end{eqnarray}
Here $\mu =\mu (t)$ and $\nu =\nu (t)$ are two arbitrary functions of $t$, $
k $ is the 3D curvature index $\left( k=\pm 1,0\right) $, and $K$ is a
constant. This solution satisfies the 5D vacuum equations $R_{AB}=0$. So we
have 
\begin{equation}
I_{1}\equiv R=0,\text{ \ }I_{2}\equiv R^{AB}R_{AB}=0,\text{ \ }
I_{3}=R_{ABCD}R^{ABCD}=\frac{72K^{2}}{A^{8}}\text{ },  \label{3-invar}
\end{equation}
which shows that $K$ determines the curvature of the 5D manifold.

Using the 4D part of the 5D metric (\ref{5-metric}) to calculate the 4D
Einstein tensor, one obtains 
\begin{eqnarray}
^{(4)}G_{0}^{0} &=&\frac{3\left( \mu ^{2}+k\right) }{A^{2}}\text{ }, 
\nonumber \\
^{(4)}G_{1}^{1} &=&^{(4)}G_{2}^{2}=^{(4)}G_{3}^{3}=\frac{2\mu \dot{\mu}}{A
\dot{A}}+\frac{\mu ^{2}+k}{A^{2}}.  \label{einstein}
\end{eqnarray}
Suppose the induced matter is a perfect fluid with density $\rho $ and
pressure $p$ moving with a 4-velocity $u^{\alpha }\equiv dx^{\alpha }/ds,$
i.e.,

\begin{equation}
^{(4)}T_{\alpha \beta }=\left( \rho +p\right) u_{\alpha }u_{\beta
}-pg_{\alpha \beta }.  \label{energy}
\end{equation}
So $u^{\alpha }=(u^{0},0,0,0)$ and $u^{0}u_{0}=1$. Substituting (\ref
{einstein}) and (\ref{energy}) into the 4D Einstein equations $^{\left(
4\right) }G_{\alpha \beta }=^{\left( 4\right) }T_{\alpha \beta }$, we find
that 
\begin{eqnarray}
\rho &=&\frac{3\left( \mu ^{2}+k\right) }{A^{2}},  \nonumber \\
p &=&-\frac{2\mu \dot{\mu}}{A\dot{A}}-\frac{\mu ^{2}+k}{A^{2}}\text{ }.
\label{dens-pres}
\end{eqnarray}

The solutions given in equations (\ref{5-metric})-(\ref{dens-pres}) contain
two arbitrary functions $\mu \left( t\right) $ and $\nu \left( t\right) $
and are, therefore, quite general. As soon as the two functions $\mu \left(
t\right) $ and $\nu \left( t\right) $ are given, the solutions are fixed. In
another hand, if the coordinate $t$ and the equation of state $p=p\left(
\rho \right) $ are fixed, we can also fix the solution. In this letter, to
illustrate the bounce properties explicitly, we will use the former to fix
the solution. That is, we let

\begin{eqnarray}
k &=&0,\text{ \ \ \ }K=1,  \nonumber \\
\nu \left( t\right) &=&t_{c}\left/ t\right. ,\text{ \ \ }\mu \left( t\right)
=t^{-1\left/ 2\right. },  \label{functions}
\end{eqnarray}
where $t_{c}$ is a constant. Substituting (\ref{functions}) into (\ref{A-B})
and (\ref{dens-pres}), we have

\begin{eqnarray}
A^{2} &=&t\left[ 1+\left( \left( y+t_{c}\right) \left/ t\right. \right) ^{2} 
\right]  \nonumber \\
B^{2} &=&\frac{1}{4}\left[ 1-\left( \left( y+t_{c}\right) \left/ t\right.
\right) ^{2}\right] ^{2}\left[ 1+\left( \left( y+t_{c}\right) \left/
t\right. \right) ^{2}\right] ^{-1}  \label{f-A-B}
\end{eqnarray}
and

\begin{eqnarray}
\rho  &=&\frac{3}{t^{2}\left[ 1+\left( \left( y+t_{c}\right) \left/ t\right.
\right) ^{2}\right] }  \nonumber \\
p &=&\frac{2}{t^{2}\left[ 1-\left( \left( y+t_{c}\right) \left/ t\right.
\right) ^{2}\right] }-\frac{1}{t^{2}\left[ 1+\left( \left( y+t_{c}\right)
\left/ t\right. \right) ^{2}\right] }.  \label{f-dens-pres}
\end{eqnarray}
Equations (\ref{f-A-B}) and (\ref{f-dens-pres}) constitute a simple exact
solution. From (\ref{f-A-B}) we can show that in a given $y=constant$
hypersurface the scale factor $A\left( t,y\right) $ has a minimum point at 
\begin{equation}
t=\left| y+t_{c}\right| \equiv t_{b}\text{ ,}
\end{equation}
at which we have 
\begin{equation}
A\left| _{_{t=t_{b}}}\right. =\left( 2t_{b}\right) ^{1\left/ 2\right. }\text{
, \ }B\left| _{_{t=t_{b}}}\right. =0\text{ , \ \ }\dot{A}\left|
_{_{t=t_{b}}}\right. =0\text{ .}
\end{equation}
So at the bounce point $t=t_{b}$ the three invariants in equation (\ref
{3-invar}) are normal. It means that there is no space-time singularity in
the big bounce model. In equation (\ref{f-dens-pres}), we can see that at
the bounce point $t=t_{b}$ the pressure undergoes a transition from negative
infinity to positive infinity, which corresponds to a phase transition of
the matter, i.e., a matter singularity. For a radially moving photon we have 
$ds^{2}=0$, so $\left( dr\left/ dt\right. \right) \left| _{t=t_{b}}\right. =0
$. This implies that $B=0$ corresponds to an event horizon. For
illustration, we plot the 3D graph of the evolution of the scale factor $
A\left( t,y\right) $ in Figure 1. 
\begin{figure}[tbp]
\centering\includegraphics[width=2.5in,height=2.5in]{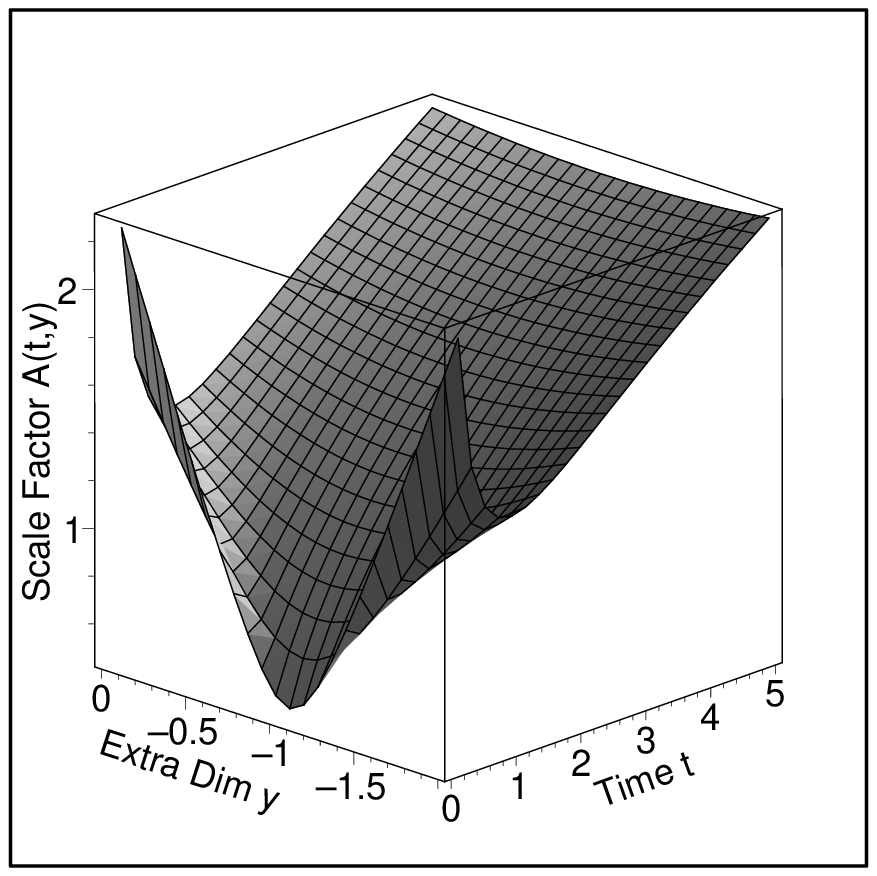}
\caption{Evolution of the scale factor $A\left( t,y\right) =\protect\sqrt{t
\left[ 1+\left( \left( y+t_{c}\right) \left/ t\right. \right) ^{2}\right] }$
with $t_{c}=1$.}
\end{figure}
From Figure 1 we see that according to the t-coordinate, the universe
evolves smoothly across its minimum at $t=t_{b}$. This strongly suggests
that time, and the arrow of time, exist before the big bounce. However, we
notice that $t$ is not the proper-time. To make sure, we need a coordinate
transformation from $t$ to the proper-time $\tau $. Now we let $
t=t_{b}=\left| y+t_{c}\right| $ corresponds to $\tau =0$, and we let the
arrow of the $\tau $-coordinate points in the same direction as the $t$
-coordinate. Then from (\ref{5-metric}) and (\ref{f-A-B}), the coordinate
transformation reads

\begin{equation}
\int_{0}^{\tau }d\tau =\int_{t_{b}}^{t}\left| B\right| dt=\frac{1}{2}
\int_{t_{b}}^{t}\left( \left| 1-\left( t_{b}\left/ t\right. \right)
^{2}\right| \cdot \left( 1+\left( t_{b}\left/ t\right. \right) ^{2}\right)
^{-1\left/ 2\right. }\right) dt.  \label{integ}
\end{equation}
The integration of (\ref{integ}) gives 
\begin{equation}
\tau \left( t\right) =\frac{1}{2}\times \left\{ 
\begin{array}{c}
-\sqrt{t^{2}+t_{b}^{2}}+t_{b}\ln \left( \frac{t}{t_{b}+\sqrt{t^{2}+t_{b}^{2}}
}\right) +C\text{ \ },\text{ \ for }0<t\leqslant t_{b} \\ 
\sqrt{t^{2}+t_{b}^{2}}-t_{b}\ln \left( \frac{t}{t_{b}+\sqrt{t^{2}+t_{b}^{2}}}
\right) -C\text{ \ },\text{ \ \ \ \ for }t_{b}\leq t\leq +\infty
\end{array}
\right. ,  \label{transfor}
\end{equation}
where

\begin{equation}
C=t_{b}\left[ \sqrt{2}-\ln \left( \sqrt{2}-1\right) \right] .  \nonumber
\end{equation}

In the coordinate transformation (\ref{transfor}), we find that there is an\
one-to-one correspondence between $t$ and $\tau $, and as $t$ varies from
zero to infinity, the proper-time $\tau $ varies from negative infinity to
positive infinity. We also find that 
\begin{equation}
\lim\limits_{t\rightarrow t_{b}^{-}}\frac{d\tau }{dt}=\lim_{t\rightarrow
t_{b}^{+}}\frac{d\tau }{dt}=0.  \nonumber
\end{equation}
It means that the proper-time joints together in a smooth way at the bounce
point. The transformation (\ref{transfor}) is shown in Figure 2 in which we
have set $t_{b}=1$ without loss of generality. 
\begin{figure}[tbp]
\centering\includegraphics[width=2.5in,height=2.5in]{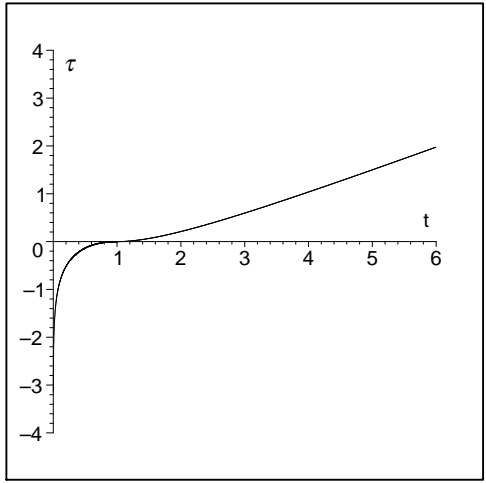}
\caption{Coordinate transformation between the coordinate-time $t$ and the
proper-time $\protect\tau $.}
\end{figure}
\ $\ \ \ $Now we discuss the evolution of the scale factor $A$ versus the
proper-time $\tau $. For simplicity, we consider it in an approximate way as
follows. For $0<t\ll t_{b}$ (corresponding to $-\infty <\tau \ll 0$), we
keep only the leading term in (\ref{transfor}), so we get

\begin{eqnarray}
2\tau &\sim &t_{b}\ln t  \nonumber \\
A &\sim &t_{b}t^{-\frac{1}{2}}\sim t_{b}e^{-\tau \left/ t_{b}\right. }
\label{as-tau}
\end{eqnarray}
Now the 5D line element of (\ref{5-metric}) reads

\begin{equation}
dS^{2}\approx d\tau ^{2}-t_{b}^{2}e^{-2\tau \left/ t_{b}\right. }\left(
dr^{2}+r^{2}d\Omega ^{2}\right) -dy^{2}\text{ .}  \label{as-5-metric}
\end{equation}
The 4D part of equation (\ref{as-5-metric}) is in fact the de Sitter metric,
which would be interpreted as having $\rho =0$ and $\Lambda =3\left/
t_{b}^{2}\right. $. In equation (\ref{as-5-metric}), the scale factor is an
exponential function of proper-time $\tau $ and corresponds to a
deflationary stage of the universe. In equation (\ref{f-A-B}), let $
t\rightarrow \infty $ (corresponding to $\tau \rightarrow \infty $), then $
A\rightarrow t^{1\left/ 2\right. },$ and $B\rightarrow 1$, the universe
expands as in the standard Friedmann-Robertson-Walker (FRW) model for the
radiation dominated epoch. At $t=t_{b}$, the scale factor reaches to its
minimum point at $A=\left( 2t_{b}\right) ^{1\left/ 2\right. }$ which
corresponds to the bounce point. Using (\ref{f-A-B}), (\ref{f-dens-pres})
and (\ref{integ}) we can show that

\begin{eqnarray}
\lim_{\tau \rightarrow 0^{-}}\frac{dA}{d\tau } &=&-\frac{1}{\sqrt{t_{b}}} 
\text{, \ \ }\lim_{\tau \rightarrow 0^{+}}\frac{dA}{d\tau }=\frac{1}{\sqrt{
t_{b}}}\text{,}  \nonumber \\
\lim_{\tau \rightarrow 0^{-}}p &=&-\infty \text{, \ \ \ }\lim_{\tau
\rightarrow 0^{+}}p=+\infty \text{ .}  \label{limit}
\end{eqnarray}

So the scale factor $A$ expressed in terms of the proper-time $\tau $ is
continuous but not smooth at bounce point. Meanwhile, the pressure undergoes
a transition from negative infinity to positive infinity. This implies that
a matter singularity exists at the bounce point. When $t\rightarrow 0$, $
\tau \rightarrow -\infty $, which means that according to the proper-time $
\tau $, the universe has existed for an infinitely long time. Consequently,
with the time elapsing in the range $\left( 0,t_{b}\right) $, the universe
contracts and crunches driven by negative pressure. At $t=t_{b}$ the
universe gets to its minimum point, and then bounces off driven by positive
pressure with radiation and matter creation. At that time the universe has a
finite density $\rho =3/\left( 2t_{b}^{2}\right) .\ $From $t_{b}$ to now,
the universe expands. In summary, the bounce singularity of a simple 5D
cosmological model is studied. We point out that the bounce singularity with 
$A\neq 0$ and $B\equiv \dot{A}\left/ \mu \right. =0$ is not a space-time
singularity. It is just a phase transition from de Sitter space to a FRW
space. At the bounce point, the scale factor $A$ is continuous but not
smooth with respect to the proper-time $\tau $, and the pressure has a jump
from negative infinity to positive infinity which corresponds to a matter
singularity and may represent a phase transition as in the inflationary
models \cite{14}. We point out that the $B=0$ singularity is an event
horizon as is in the Schwarzschild solution. According to the proper-time,
the whole bounce scenario can be described as follows. Our universe has been
existed for an infinitely long time. Before the bounce, the universe
contracts deflationary. When it approaches to the bounce point, it undergoes
a crunch driven by an infinite negative pressure, and then it bounces off
driven by an infinite positive pressure and accompanied by creation of
radiation and matter. After the bounce, the universe expands asymptotically
as is in the standard FRW models.

\textbf{Acknowledgements} We thank Paul Wesson and Guowen Peng for comments.
This work was supported by NSF of P.R. China under Grants 19975007 and
10273004.

\end{document}